\documentclass[prb,twocolumn,superscriptaddress,showpacs,amsmath,amssymb]{revtex4-2}
\usepackage[colorlinks=true, linkcolor=blue, filecolor=blue, urlcolor=blue, citecolor=blue]{hyperref}
\usepackage{amsfonts}
\usepackage{bbm}
\usepackage{graphicx}% Include figure files
\usepackage{dcolumn}% Align table columns on decimal point
\usepackage{bm}% bold math
\usepackage{ulem}

\begin{document}
	
	\title{Tunneling Magnetoresistance Effect in Altermagnets}
	
	\author{Yu-Fei Sun}
	\affiliation{International Center for Quantum Materials, School of Physics, Peking University, Beijing 100871, China}
	\affiliation{Hefei National Laboratory, Hefei 230088, China}

	\author{Yue Mao}
	\affiliation{International Center for Quantum Materials, School of Physics, Peking University, Beijing 100871, China}

	\author{Yu-Chen Zhuang}
        \email[]{yuchenzhuang@pku.edu.cn}
	\affiliation{International Center for Quantum Materials, School of Physics, Peking University, Beijing 100871, China}
	
	\author{Qing-Feng Sun}
	\email[]{sunqf@pku.edu.cn}
	\affiliation{International Center for Quantum Materials, School of Physics, Peking University, Beijing 100871, China}
	\affiliation{Hefei National Laboratory, Hefei 230088, China}
	\affiliation{Collaborative Innovation Center of Quantum Matter, Beijing 100871, China}

\begin{abstract}
As an unconventional magnet, altermagnetism attracts great interest in condensed matter physics and applies a new research platform for the spintronics. Since the tunneling magnetoresistance (TMR) effect is an important research aspect in spintronics, we theoretically propose a universal altermagnetic sandwich device to achieve the TMR effect and investigate its transport properties. Using the nonequilibrium Green’s function method and the Landauer-B\"uttiker formula, we obtain the conductance and the TMR ratio. By systematically rotating the orientations of the altermagnet and spin, we investigate how the altermagnetic orientations affect the conductance and the TMR ratio, and comprehensively demonstrate the dependence of the conductance and the TMR ratio on a range of parameters in the system. By tuning the altermagnetism strength and the Fermi energy, as well as rotating the orientations in the altermagnet, the TMR ratio can reach a value of over 1000\%. In addition, we analyze the detailed symmetry relations of the conductance and the TMR ratio in our system. Our approach provides a new design concept for the next-generation information technologies based on the altermagnetic platform, paving the way for the development of spintronics applications.
\end{abstract}
\maketitle
	
\section{\label{sec1}Introduction}
Magnetism, as one of the oldest disciplines in physics, has remained a vibrant and evolving field of research. Its rich phenomena have not only deepened the understanding of physics, but also played a central role in technological applications ranging from data storage to magnetic sensing. Traditionally, magnetic materials have been broadly classified into ferromagnet (FM) and antiferromagnet (AFM) \cite{smejkal_emerging_2022,smejkal_conventional_2022,lee_broken_2024,krempasky_altermagnetic_2024}. FMs are characterized by a spontaneous net magnetization, and they have already underpinned the majority of magnetic technologies due to their strong spin polarization and ease of control via external magnetic fields. In contrast, AFMs have different symmetry-related sublattice with FMs and are characterized by a vanishing net magnetization \cite{samanta_spin_2025,liu_giant_2024}. 
The early works mainly focused on conventional AFMs with collinear spin configurations. Although these conventional AFMs have rich and intriguing physics, they have found limited practical applications due to their challenges in spin detection and manipulation \cite{liu_different_2025,shao_spinneutral_2021}.
%Historically, early works focused on AFMs concentrated with collinear spin configurations, which is referred to as conventional AFMs nowadays. Although these conventional AFMs have fancy and intrinsic physics, they have found less practical application due to their difficulties in spin detection and manipulation.

However, with the advance of antiferromagnetic spintronics and the development of advanced experimental techniques, researchers have found some unconventional magnets within the category of AFMs \cite{liu_different_2025}, such as some non-collinear antiferromagnets \cite{chen_anomalous_2014a,sticht_noncollinear_1989a,chen_octupoledriven_2023,qin_roomtemperature_2023a,zelezny_spin_2018,zelezny_spinpolarized_2017}, which may exhibit both antiferromagnetic and ferromagnetic properties. These unconventional magnets could combine the advantages of AFM and FM that can host spin-polarized currents and also enable ultrafast spin manipulation with low stray field. This has opened new avenues for spintronics applications and attracted significant interest \cite{shao_spinneutral_2021}. 

Recently, researchers have also identified one prominent example of unconventional magnets, named altermagnet (AM) \cite{smejkal_conventional_2022,smejkal_emerging_2022,cheng_fieldfree_2024a,fukaya_josephson_2025,hoyer_spontaneous_2025,lee_broken_2024,mazin_editorial_2022,PhysRevB.111.045407,rooj_altermagnetism_2025,shao_spinneutral_2021,smejkal_giant_2022,smolyanyuk_origin_2025,sun_andreev_2023,PhysRevB.111.035423,PhysRevLett.132.056701,fang_quantum_2024,PhysRevB.108.L060508,bose_tilted_2022,bose_tilted_2022,zhou_manipulation_2025a,PhysRevB.110.155125,PhysRevB.110.094508,PhysRevB.111.064422,PhysRevB.110.L060508,PhysRevB.111.064428}.
Although AM generally has a collinear antiferromagnetic magnetic order, it additionally exhibits special rotation symmetry characteristics in space, resulting in a ferromagnetic-like spin-splitting of the anisotropic band structure. 
Based on non-relativistic spin-symmetry groups, some researchers have proposed that AMs could constitute a third fundamental collinear magnetic phase, as discussed in their work \cite{smejkal_conventional_2022,smejkal_emerging_2022}. 
It is derived that AMs are characterized by crystal-rotation symmetries connecting opposite-spin sublattices in real space and opposite-spin energy bands in momentum space.
Within the framework of crystal-rotation symmetries, AM exhibits an adjustable angular degree of freedom (i.e., the orientation of AM) , which can exhibit intriguing anisotropic spin-transport properties and has attracted considerable attentions \cite{PhysRevLett.130.216701,gonzalez-hernandez_efficient_2021,cheng_orientationdependent_2024,liao_separation_2024}.

Using techniques such as synchrotron-based angle-resolved photoemission spectroscopy (ARPES) and spin-resolved ARPES, experimental studies have verified altermagnetism in various magnetic materials like CrSb, MnTe, etc \cite{zhou_manipulation_2025a,yang_threedimensional_2025,lee_broken_2024,krempasky_altermagnetic_2024, PhysRevB.109.115102,zhu_observation_2024,Ole_Observat_2024,jiang_metallic_2025a}. In addition, through electrical, magnetic and other methods, experiments have been able to adjust the orientation of the magnetic-order (N\'eel) vector of AM, and further discovered a series of intriguing physical phenomena and offered numerous potential applications, such as anomalous Hall effects \cite{PhysRevLett.130.036702,wang_emergent_2023a,PhysRevB.109.224430,feng_anomalous_2022} and spin-torque effects \cite{han_electrical_2024,PhysRevLett.128.197202,PhysRevLett.129.137201}. These findings expand the understanding of magnetic materials and provide possibilities to develop novel applications in spintronics and next-generation information technologies.

A key focus in spintronics research is the tunneling magnetoresistance (TMR) effect in magnetic materials \cite{JULLIERE1975225,PhysRevLett.74.3273,Yuasa_2007}, which provides a groundbreaking mechanism for manipulating the conductance and plays a crucial role in data storage and sensing technologies.
TMR effect occurs in the magnetic tunnel junction, a sandwich structure composed of two ferromagnetic electrodes separated by a thin insulating tunnel barrier, where electron transport occurs via spin-dependent quantum tunneling.
This phenomenon focuses on the change in electrical conductance depending on the relative magnetization alignment between the two magnetic electrodes. As the magnetization of the two magnetic electrodes can be controlled antiparallelly or parallelly, the junction can be switched between high and low resistance, forming the core of the magnetic random-access memories (MRAMs), which is one of the most well-known applications of spintronics \cite{yang_twodimensional_2022,han_spinorbit_2021}.
While FM-based TMR devices have been widely adopted in industrial applications, AFM-based systems have attracted growing interest due to their potential for ultrafast spin dynamics and inherent robustness against external magnetic perturbations \cite{samanta_spin_2025,jungwirth_antiferromagnetic_2016}.
However, the lack of global spin polarization has traditionally limited the use of AFM as electrodes in TMR devices \cite{chi_crystalfacetoriented_2024}.
Consequently, the exploration of the unconventional magnets AM may become an intriguing subject, as it incorporates features of FM and AFM \cite{smejkal_emerging_2022,yang_threedimensional_2025}. 
We wonder whether there exist richer and more versatile TMR effects based on AM.

The existing experimental studies have provided a solid foundation and promising avenues for exploring TMR in AM \cite{chi_crystalfacetoriented_2024,PhysRevLett.129.137201,PhysRevLett.128.197202,PhysRevLett.130.036702,feng_anomalous_2022,wang_emergent_2023a,PhysRevB.109.224430}.
It is also noted that a few theoretical studies by first-principles calculations \cite{yan_giant_2024,liu_giant_2024,chi_crystalfacetoriented_2024} and very recently an experiment \cite{samanta_spin_2025} have indicated that the TMR effect can be found along
certain specific crystallographic orientations of AM.
However, these researches only explore TMR along certain orientations. There is still a lack of systematic theoretical study to illuminate the properties of AM-based TMR, especially on how the various orientations in AM influence the transport and TMR.

\begin{figure}[!htb]
\centerline{\includegraphics[width=\columnwidth]{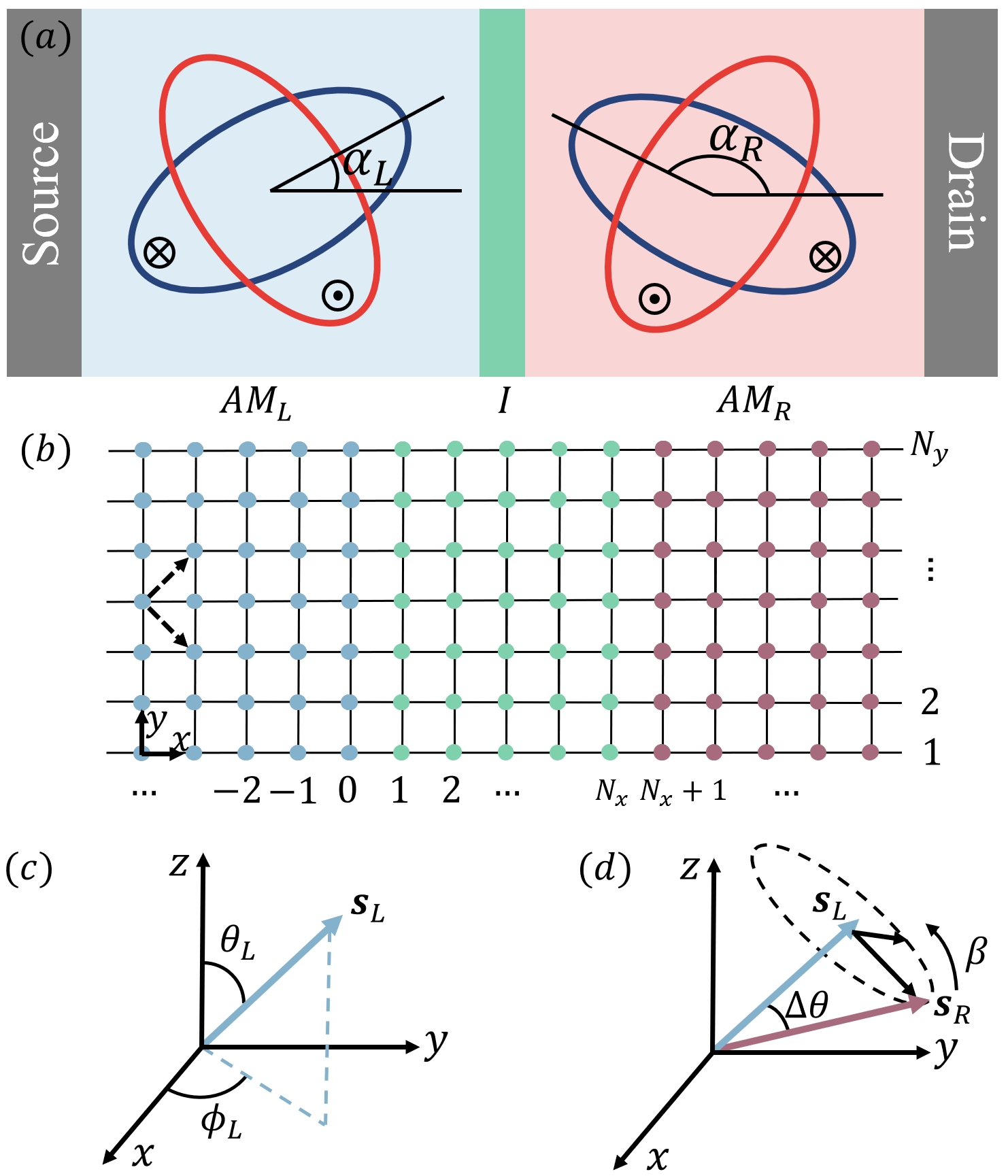}}
    \caption{
(a) Schematic of the device: two altermagnetic materials are placed on the left and right sides of the insulator, and the source and drain leads are coupled to this AM-I-AM junction. The two ellipses in the AM region denote the Fermi surfaces for different spins of electrons. The red one is for the spin-up state and the blue one is for the spin-down state. The orientation of AM is described by the angle $\alpha_{L(R)}$ between the crystalline axis (the major axis of the blue ellipse) and the normal direction of the interface.
(b) Schematic illustration of the two-dimensional discretized square lattice model obtained from the continuum model.
The dashed arrows represent the next-nearest-neighbor hoppings in AMs.
The lattice constant is $a$. The length of the center insulating layer is $N_x$ and the width of the junction is $N_y$.
(c) The orientation of spin for the left $\mathrm{AM}$. It can point to arbitrary directions, defined by the azimuthal angle $\phi_L$ and the polar angle $\theta_L$.
(d) The blue arrow represents the orientation of spin for the left $\mathrm{AM}$, $\mathbf{s}_L$, while the red arrow represents the orientation of spin for the right $\mathrm{AM}$, $\mathbf{s}_R$. The relative spin angle between the two spins is denoted as $\Delta \theta$. When keeping $\Delta \theta$ fixed, the orientation of $\mathbf{s}_R$ can rotate around $\mathbf{s}_L$, with the rotational phase angle $\beta$.
    }
    \label{Fig1}
\end{figure}

In this paper, we theoretically investigate a system with an insulator sandwiched between two AMs, i.e., the AM-insulator-AM (AM-I-AM) junction, as shown in Fig. \ref{Fig1}(a). The orientation of the momentum anisotropy of AM, which we call the orientation of AM for short, can be denoted by the angle $\alpha$ between the crystalline axis (the major axis of the blue ellipse) and the interface normal (transport direction).  
The unique rotational symmetry characteristic of AM indicates that the angle $\alpha$ should be an important factor to affect TMR. It also inherently refers to the crystallographic direction of AM, as the anisotropic spin splitting and band structure in momentum space are directly governed by the real-space crystal symmetry.
In addition to the orientation of AM, we also take the orientation of spin into account (which is related to the N\'eel vector). As shown in Fig. \ref{Fig1}(c), the orientation of the spin is described by the polar angle $\theta$ and the azimuthal angle $\phi$.

By applying the nonequilibrium Green's function method and Landauer-B\"uttiker formula, we calculate the conductance across the AM-I-AM junction under different orientations in AM.  Due to the anisotropy of the spin-resolved Fermi surface, the conductance will vary with the orientation of AM, leading to a substantial TMR effect. By fixing the orientation of spin for the left AM and right AM at $(\theta_L, \phi_L)= (\theta_R, \phi_R)=(0, 0)$, we first investigate the conductance and TMR ratio by rotating the orientation of the right AM $\alpha_R$ with the orientation of the left AM $\alpha_L$ fixed. Next, we investigate the TMR effect by rotating both orientations of the left and right AM. Furthermore, we fix the orientation of AM at $\alpha_L=\alpha_R=0$ and study how the orientation of spin affects the conductance and TMR ratio. Using unitary transformation, we show that the conductance and TMR ratio only depend on the relative spin angle $\Delta \theta$ between the left and right spin, as shown in Fig. \ref{Fig1}(d). Ultimately, the combined effect of both the orientations of AM and spin on the conductance and TMR ratio is demonstrated. The TMR ratio can reach over 1000$\%$ by optimizing the system parameters. 
Our work can enlighten the quantum transport and magnetoresistance effect in AM and open horizons to develop AM-based spintronics devices.

The rest of this paper is organized as follows.
In Sec. \ref{sec2}, we present the Hamiltonian for our altermagnetic sandwich system and calculate the conductance and TMR ratio by the nonequilibrium Green’s function method and the Landauer-B\"uttiker formula in the lattice model.
In Sec. \ref{sec3}, we present the numerical results of the conductance and TMR ratio for different orientations of AM with the fixed orientation of spin.
In Sec. \ref{sec4}, we present the numerical results of the conductance and TMR ratio for different orientations of spin with the fixed orientation of AM.
In Sec. \ref{sec5}, we thoroughly study the combined effect from orientations of AM and spin. Discussion and conclusion are given in Sec. \ref{sec6}.
The discretization of the tight-binding Hamiltonian is given in the Appendix \ref{A}.  In Appendix \ref{B}, we present the influence of the length of the center insulating layer on our results.

\section{\label{sec2}Model and formulations}
We consider the two-dimensional AM-I-AM junction in the $xy$ plane, as shown in Fig. \ref{Fig1}(a). The two ellipses in AM areas are Fermi surfaces occupied by electrons with spin-up state (red color) and spin-down state (blue color). The orientation of the left (right) AM is characterized by $\alpha_{L(R)}$, which is defined as the angle between the crystalline axis of the left (right) AM and the interface normal.

In the continuum model, the Hamiltonian of the left AM $H_{L}(\mathbf{\hat{k}})$, the right AM $H_{R}(\mathbf{\hat{k}})$ and the insulator $H_{I}(\mathbf{\hat{k}})$ can be written as:
\begin{eqnarray}
H_{L/R/I}(\mathbf{\hat{k}})=\sum_{\mathbf{\hat{k}}}\Psi_{{L/R/I}\mathbf{\hat{k}}}^{\dagger} \check{H}_{L/R/I}(\mathbf{\hat{k}}) \Psi_{{L/R/I}\mathbf{\hat{k}}},
\end{eqnarray}
with $\Psi_{{L/R/I}\mathbf{\hat{k}}}=(\Psi_{{L/R/I}\mathbf{\hat{k}}\uparrow },\Psi_{{L/R/I}\mathbf{\hat{k}}\downarrow})^{T}$.
$\Psi_{{L/R/I}\mathbf{\hat{k}}\uparrow (\downarrow) }$ is the annihilation operator of the electron with spin up (down).

Here, $\check{H}_{L(R)}(\mathbf{\hat{k}}) = H_0^{L(R)}(\mathbf{\hat{k}})+H_J^{L(R)}(\mathbf{\hat{k}})$ and $\check{H}_{I} (\mathbf{\hat{k}})=H_0^{I}(\mathbf{\hat{k}})$.
The term $H_0$ is expressed as:
\begin{equation}
    \begin{aligned}
&H_0^{L/R/I}(\mathbf{\hat{k}}) = [t_0\left ( \hat{k}_x^2 + \hat{k}_y^2 \right ) - \mu_{L/R/I}] \sigma_0.
    \end{aligned}
\end{equation}
The altermagnetic term $H_J$ describes a $d$-wave altermagnet, which is derived from symmetry considerations consistent with the representative spin-group classification \cite{smejkal_conventional_2022}. The evidence of similar altermagnetic splitting has also been observed in a recent experiment \cite{wei_$mathrmla_2mathrmo_3mathrmmn_2mathrmse_2$_2025}.
The methodology adopted in this paper can also be applied to various 
other altermagnets (such as g-wave or i-wave altermagnets),
as the nonequilibrium Green's function method used here has a wide range of applications. Then similar TMR effects should also be achievable.
Since $d$-wave altermagnetism is one of the most common types of altermagnetism, 
we focus on studying it here.
The $d$-wave altermagnetic term $H_J$
is expressed as \cite{cheng_orientationdependent_2024,cheng_fieldfree_2024a,sun_andreev_2023,PhysRevB.108.L060508}:
\begin{equation}
    \begin{aligned}
&H_J^{L(R)}(\mathbf{\hat{k}}) = t_J\left [ \left ( \hat{k}_x^2 - \hat{k}_y^2 \right ) \cos 2\alpha_{L(R)} +2\hat{k}_x \hat{k}_y\sin 2\alpha_{L(R)} \right ] \bm{\sigma}\cdot \mathbf{s}_{L(R)}. \\
    \end{aligned}
\end{equation}

Here, $t_0$ is the kinetic coefficient, $t_J$ denotes the altermagnetism strength, and $\mu_{L/R/I}$ are chemical potentials in the regions of the left AM, center insulator and right AM.
$\bm{\sigma}=\left( \sigma_x, \sigma_y, \sigma_z \right)$ are the Pauli matrices, and $\sigma_0$ is the identity matrix.
$\mathbf{\hat{k}}=(\hat{k}_x, \hat{k}_y)$ is the two-dimensional momentum operator with $\hat{k}_x=-i\frac{\partial}{\partial x}$, $\hat{k}_y=-i\frac{\partial}{\partial y}$, and we take $\hbar=1$.
To further consider the effect of spin-rotation,
we set $\mathbf{s}_{L(R)}=(\sin \theta_{L(R)} \cos \phi_{L(R)}, \sin \theta_{L(R)} \sin \phi_{L(R)}, \cos \theta_{L(R)})$ which denotes the orientation of spin in the left (right) AM on the Bloch sphere. $\mathbf{s}_{L(R)}$ can point at arbitrary directions by rotating the azimuthal angle $\phi_{L(R)}$ and the polar angle $\theta_{L(R)}$, as shown in Fig. \ref{Fig1}(c).

To calculate the conductance of the AM-I-AM junction, we can discretize the Hamiltonian on a two-dimensional square lattice with the lattice constant $a$, as shown in Fig. \ref{Fig1}(b).
We set the site number along the $y$ direction of the junction (i.e, width) is $N_y$ and the site number of the center insulator along the $x$ direction is $N_x$. The left (right) semi-infinite AM area is placed in the electrode region $x \le 0$ ($x \ge N_x+1$).
The whole discrete Hamiltonian can be given by \cite{PhysRevB.111.054515}:
\begin{equation}
H_{dis}=H_{L}+H_{R}+H_{I}+H_{T}.
\end{equation}
The detailed discretization of Hamiltonians for the left (right) AMs $H_{L(R)}$, the center insulator $H_I$, and the tunneling Hamiltonian $H_T$ between them is shown in the Appendix \ref{A}. Due to the specific symmetry of the altermagnetic Hamiltonian, we emphasize that the next-nearest-neighbor hopping terms $\check{H}_{xy}$ and $\check{H}_{x\bar{y}}$ will naturally arise in the discretized form [indicated by dashed arrows in Fig.~\ref{Fig1}(b)], resulting in an intrinsic momentum-dependent spin splitting of AMs.

By using the nonequilibrium Green’s function method \cite{PhysRevB.108.214519,PhysRevB.111.054515,PhysRevB.105.165148} and the Landauer-B\"uttiker formula \cite{zhang_spin_2010}, the conductance $G(E)$ is calculated as:
\begin{equation}
%\frac{e^2}{h}
G(E)=\frac{e^2}{h}Tr \left[ \mathbf{\Gamma}_L(E) \mathbf{G}^r(E)\mathbf{\Gamma}_R(E)\mathbf{G}^a(E) \right],
\end{equation}
where $\mathbf{G}^r(E)=[\mathbf{G}^a(E)]^{\dagger}=[E\mathbf{I}-\mathbf{H}_I
-\mathbf{\Sigma}_L^r(E)-\mathbf{\Sigma}_R^r(E)]^{-1}$ is the retarded Green's function, and $\mathbf{\Gamma}_{L(R)}(E)=i\left \{\mathbf{\Sigma}_{L(R)}^r(E)-\left[ \mathbf{\Sigma}_{L(R)}^r(E) \right]^\dagger \right \}$ is the linewidth function. $\mathbf{\Sigma}_{L(R)}^r(E)$ is the self-energy calculated from the semi-infinite left (right) AM \cite{PhysRevB.107.184511,PhysRevB.109.045430,PhysRevB.23.4997,M_P_Lopez}. $E$ is the Fermi energy.

For two distinct magnetic configurations of AM-I-AM junctions (e.g., parallel spin and anti-parallel spin, typically), we can calculate the corresponding conductance $G_{1,2}$, respectively. Then, we can further estimate the TMR effect by the TMR ratio, which is defined as \cite{yuasa_giant_2004}:
\begin{equation}
    TMR=\frac{|G_2-G_1|}{\min(G_2, G_1)}. \label{TMR}
\end{equation}

In the following numerical calculations, unless explicitly stated, we take $t_0=1$, $N_y=40$ and $N_x=5$. The lattice constant is $a=1$. The altermagnetism strength is $t_J=0.5$. The coupling energy between different regions is $t=-1$ (details in Appendix \ref{A}). For the chemical potential, we take $\mu_L=\mu_R=0$ and $\mu_I=-1$.
Then the center region is insulating when $E+\mu_I <0$.
Units of the conductance and energy are chosen with $\frac{e^2}{h}$ and $\frac{t_0}{a^2}$, respectively. The calculation is performed under ideal conditions, in the absence of disorder.

\section{\label{sec3}The dependence of TMR effect on the orientation of AM}

In this section, we will study how the orientation of AM $\alpha$ influences the conductance and the TMR ratio with the fixed orientations of spins $\mathbf{s}_{L/R}$. In Sec.~\ref{3A}, we will focus on the conductance and TMR ratio for two typical magnetic configurations ($\alpha_{L}=\alpha_{R}=0$) and ($\alpha_{L}=0,\alpha_{R}=0.5\pi$).
Next, we demonstrate the conductances and the TMR ratios by freely rotating the orientation of the right AM $\alpha_{R}$ under the fixed orientation of the left AM as $\alpha_{L}=0$ in Sec. \ref{sec3B}.
Last, we discuss the conductance and the TMR ratios by rotating both the orientations of the left and right AM in Sec. \ref{3C}. Note that the orientations of spins $\mathbf{s}_{L/R}$ in the left and right AMs are always fixed at $(\theta_L, \phi_L)=(\theta_R, \phi_R)=(0, 0)$ in this section.

\begin{figure*}[!htb]
\centerline{\includegraphics[width=1.5\columnwidth]{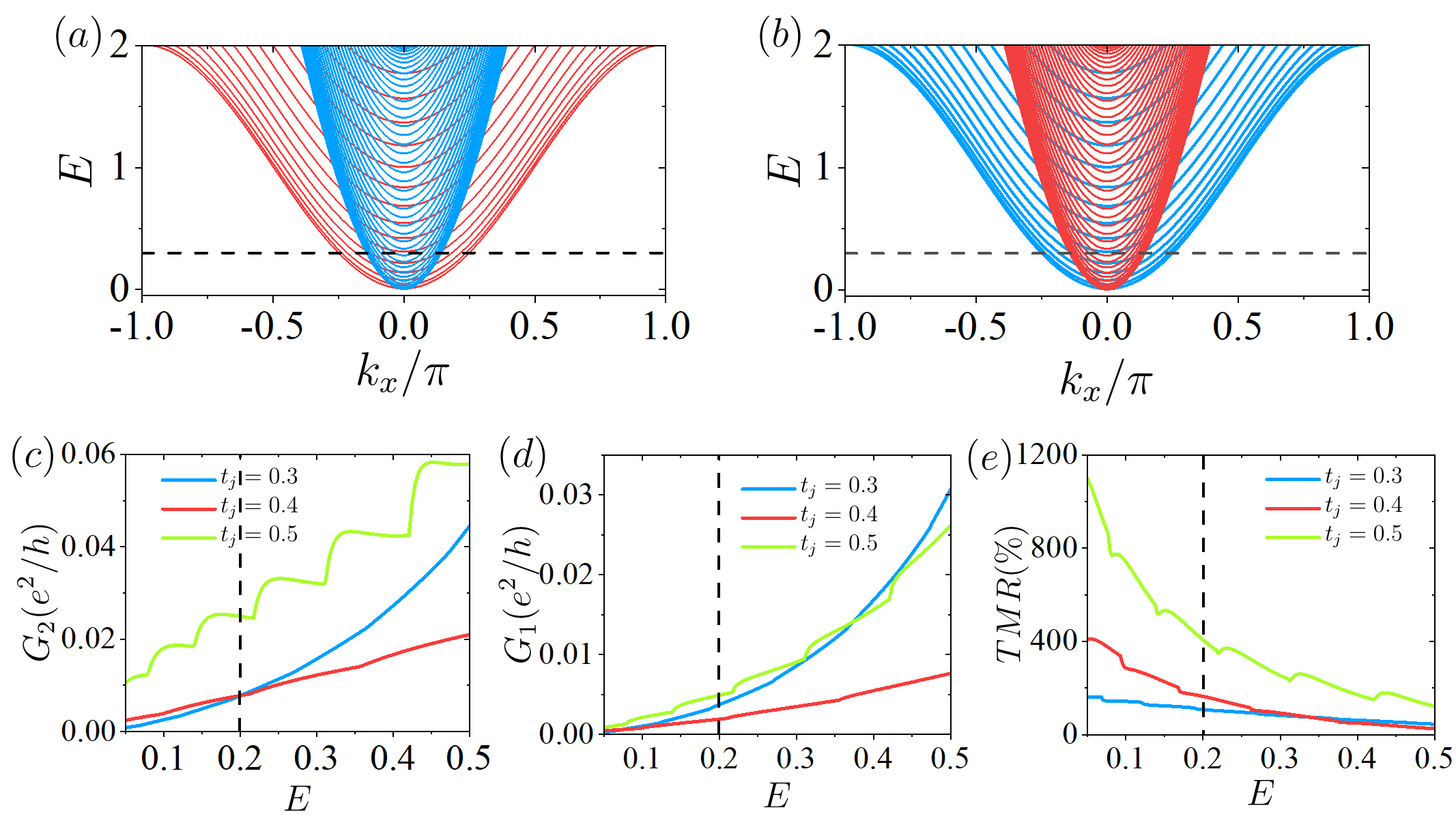}}
    \caption{
(a, b) The energy band structures of the AM nanoribbon along the $x$ direction with two typical orientations with (a) $\alpha=0.5\pi$ and (b) $\alpha=0$.
The red color curves denote the spin-up states while the blue color curves denote the spin-down states. The altermagnetism strength is $t_J=0.5$.
(c, d) are the conductance versus the Fermi energy $E$ with different $t_J$. In (c), the orientations of the left and right AMs are $\alpha_L=\alpha_R=0$, where the conductance is denoted as $G_2$. In (d), the orientations of the left and right AMs are $\alpha_L=0,\alpha_R=0.5\pi$ and the conductance is denoted as $G_1$.
(e) The TMR ratio defined in Eq. (\ref{TMR}) versus the Fermi energy $E$ with different $t_J$. The black dashed line marks $E=0.2$.}
    \label{Fig2}
\end{figure*}

\subsection{\label{3A}Appearance of TMR effect}

In Figs. \ref{Fig2}(a) and \ref{Fig2}(b), we present the energy bands of the AM nanoribbon along the $x$ direction for two typical orientations of AM, respectively [$\alpha=0.5 \pi$ for (a) and $\alpha=0$ for (b)]. Due to the combined effects of broken time-reversal symmetry and preserved crystal rotation symmetries in AM, spin splitting arises in the AM, leading to momentum-dependent spin polarization.
The red color curves in Figs. \ref{Fig2}(a) and \ref{Fig2}(b) denote the spin-up states, while the blue color curves denote the spin-down states.
It can be seen that the spin bands of AMs with orientations $\alpha=0$ and $\alpha=0.5\pi$ are just reversed.
This can be understood since the two elliptical Fermi surfaces with opposite spins are just mutually perpendicular, as shown in Fig. \ref{Fig1}(a).

The conductance from the left electrode to the right electrode depends on the degree of the matching between the incident and outgoing spin channels. When the magnetic configuration is $\alpha_L=\alpha_R=0$, the energy bands of the left and right AMs match well. Electrons are easy to tunnel through the insulating region, resulting in a higher conductance, denoted as $G_2 = G(\alpha_{L}=0, \alpha_R=0)$ [see Fig. \ref{Fig2}(c)]. Conversely, as the magnetic configuration is $\alpha_L=0$ and $\alpha_R=0.5\pi$, the spin alignments between the left and right AMs are completely opposite. A critical mismatch significantly suppresses the tunneling probability and thus the conductance is much lower, which is denoted as $G_1 = G(\alpha_{L}=0, \alpha_R=0.5 \pi)$, as shown in Fig. \ref{Fig2}(d).

The difference between $G_{1}$ and $G_{2}$ clearly indicates a TMR effect in the AM-I-AM junction. As shown in Fig. \ref{Fig2}(e), we demonstrate the TMR ratio. As the Fermi energy $E$ declines and approaches the bottom of the energy band $E=0$, we can find a significant climb in the TMR ratio, although the conductances decrease [see Figs. \ref{Fig2}(c) and \ref{Fig2}(d)].
In addition, we present the results for different values of $t_J$. The altermagnetism strength $t_J$ directly determines the magnitude of spin splitting in the AM and significantly affects both the conductance and the TMR ratio.
As $t_J$ increases, the spin polarization becomes larger, leading to an enhancement of the TMR ratio.
As a result, by fine-tuning $t_J$ and the Fermi energy $E$, the TMR ratio can reach beyond 1000$\%$, which can improve the sensitivity of the TMR sensor and enhance its application performance \cite{s25061730,AN2025116174}.
Furthermore, the TMR ratio can also serve as physically meaningful indicators to qualitatively indicate the degree of spin polarization. A larger altermagnetism strength $t_J$ leads to the stronger spin splitting in the momentum space at the same Fermi level, resulting a higher TMR ratio and a greater spin-dependent asymmetry in transmission. 
In addition, as $t_J \rightarrow 0$, the altermagnetic term $H_J$ will vanish, and the total Hamiltonian of our system is just reduced to a trivial $H_0$ with spin-degenerate bands. In this case, the system becomes a nonmagnetic ordinary metal, so no spin-related characteristics exists and the TMR effect will disappear.
The discussion about how the length of the center insulating layer affects the TMR ratio is also shown in Appendix \ref{B}. Within an appropriate range of length, it will not affect the value of the TMR ratio.
For clarity, in the following discussion, the Fermi energy is set at $E=0.2$ [the black dashed line in Figs. \ref{Fig2}(c-e)] and the length of the center insulating layer is fixed at $N_x=5$.

\subsection{\label{sec3B} Regulation of $\alpha_R$ on TMR effect}
Due to the strong anisotropy of the energy bands for AM, when varying the orientation of AM, the conductance through the junction will be dramatically affected. In this subsection, we fix the orientation of the left AM at $\alpha_L=0$, freely rotate the orientation of the right AM from $\alpha_R=0$ to $\alpha_R=\pi$, and investigate the TMR effect.

\begin{figure}[!htb]
\centerline{\includegraphics[width=\columnwidth]{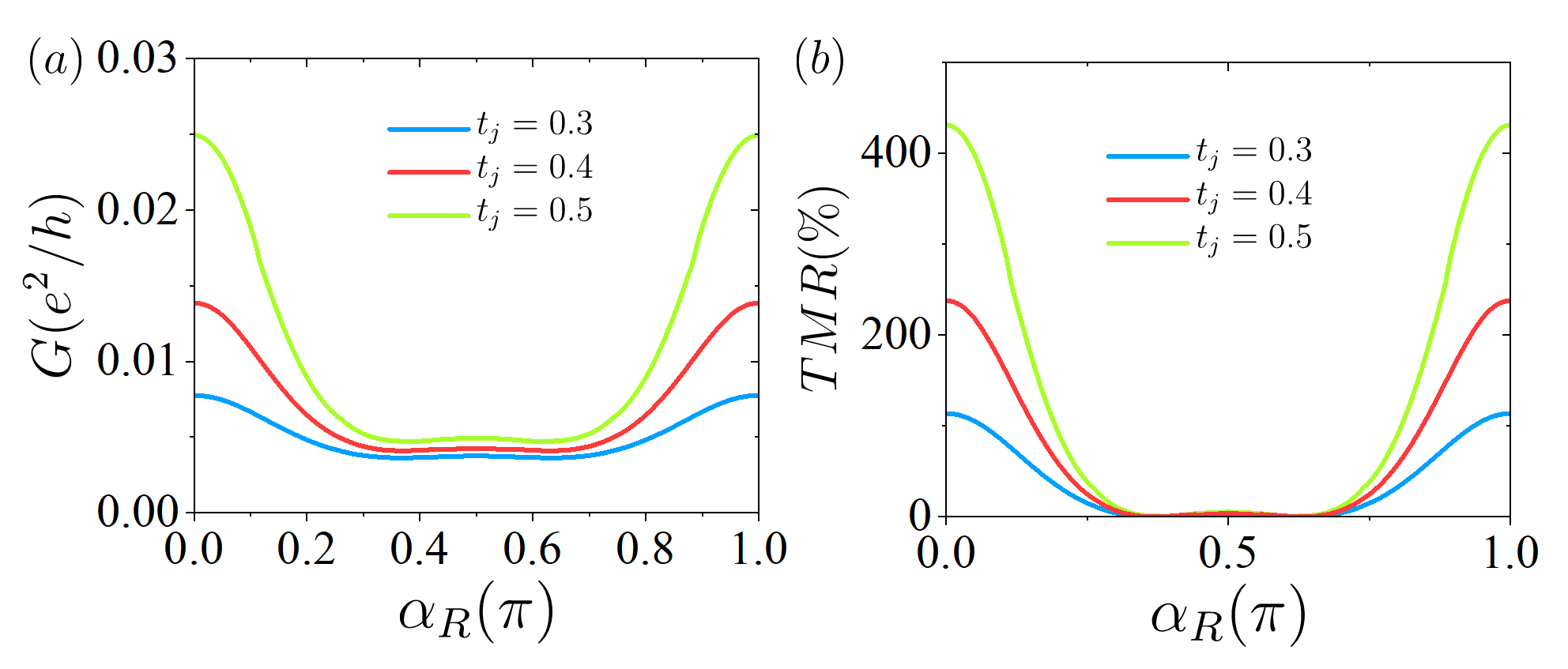}}
    \caption{
(a) The conductance versus the orientation of the right AM $\alpha_R$ under different altermagnetism strength $t_J$.  The orientation of the left AM is fixed at $\alpha_L=0$. (b) The TMR ratio versus the orientation of the right AM $\alpha_R$ under different altermagnetism strength $t_J$. Here the conductance at each value of $\alpha_R$ is denoted as $G_2 = G(\alpha_L=0, \alpha_R)$. The minimum value among them is specifically chosen as $ G_1 = \min[G(\alpha_L=0, \alpha_R)]$. The Fermi energy is set at $E=0.2$.
}
    \label{Fig3}
\end{figure}

In Fig. \ref{Fig3}(a), we plot the conductance versus the orientation of the right AM $\alpha_R$ under different $t_J$.
When $\alpha_R$ rotates from 0 to $0.5\pi$, the conductance $G$ roughly has a decreasing trend, since the mismatch between the left AM and the right AM becomes larger. As $\alpha_R$ rotates from $0.5\pi$ to $\pi$, the mismatch between the left AM and the right AM gradually weakens, and the conductance $G$ roughly has an increasing trend. In particular, the conductance $G(\alpha_{L}=0, \alpha_R)$ is symmetric about $\alpha_R=0.5\pi$.
It can be explained as follows.
Since the transport in this system occurs along the $x$-direction, the transmission probability depends on the matching of the $x$-axis projection of the energy bands on the left and right AMs. When the left AM energy band remains unchanged, and the right AM energy band has identical $x$-axis projections in orientation $\alpha_R$ and $\pi-\alpha_R$, the transmission probability is the same in these two orientations. Consequently, the conductance $G(\alpha_R)=G(\pi-\alpha_R)$ holds in this system.

In Fig.~\ref{Fig3}(b), we plot the TMR ratio versus the orientation of the right AM $\alpha_R$ under different $t_J$. For clarity, here we choose the minimum conductance among the data in Fig.~\ref{Fig3}(a) as $G_1=\min[G(\alpha_{L}=0,\alpha_{R})]$. And $G_2 = G(\alpha_{L}=0, \alpha_R)$ is the conductance with any arbitrary $\alpha_R$. 
Then TMR ratio can be obtained from Eq.(\ref{TMR}) and is shown in Fig.~\ref{Fig3}(b).
The trend of the curves is the same as that in Fig.~\ref{Fig3}(a) and the TMR ratio reaches its maximum at $\alpha_R=0, \pi$, while it is very small near
$\alpha_R=0.5\pi$. In addition, the TMR ratio also exhibits a relation: $TMR(\alpha_{L}=0,\alpha_R)=TMR(\alpha_{L}=0,\pi-\alpha_R)$.

\subsection{\label{3C}Regulation of arbitrary orientation on TMR effect}

In Sec. \ref{sec3B}, we fix the orientation of the left AM, and study the influence of the orientation of the right AM on the conductance and the TMR ratio.
In this subsection, we rotate both the orientation of the left and right AM and study the combined effect of their orientations.

\begin{figure}[!htb]
\centerline{\includegraphics[width=\columnwidth]{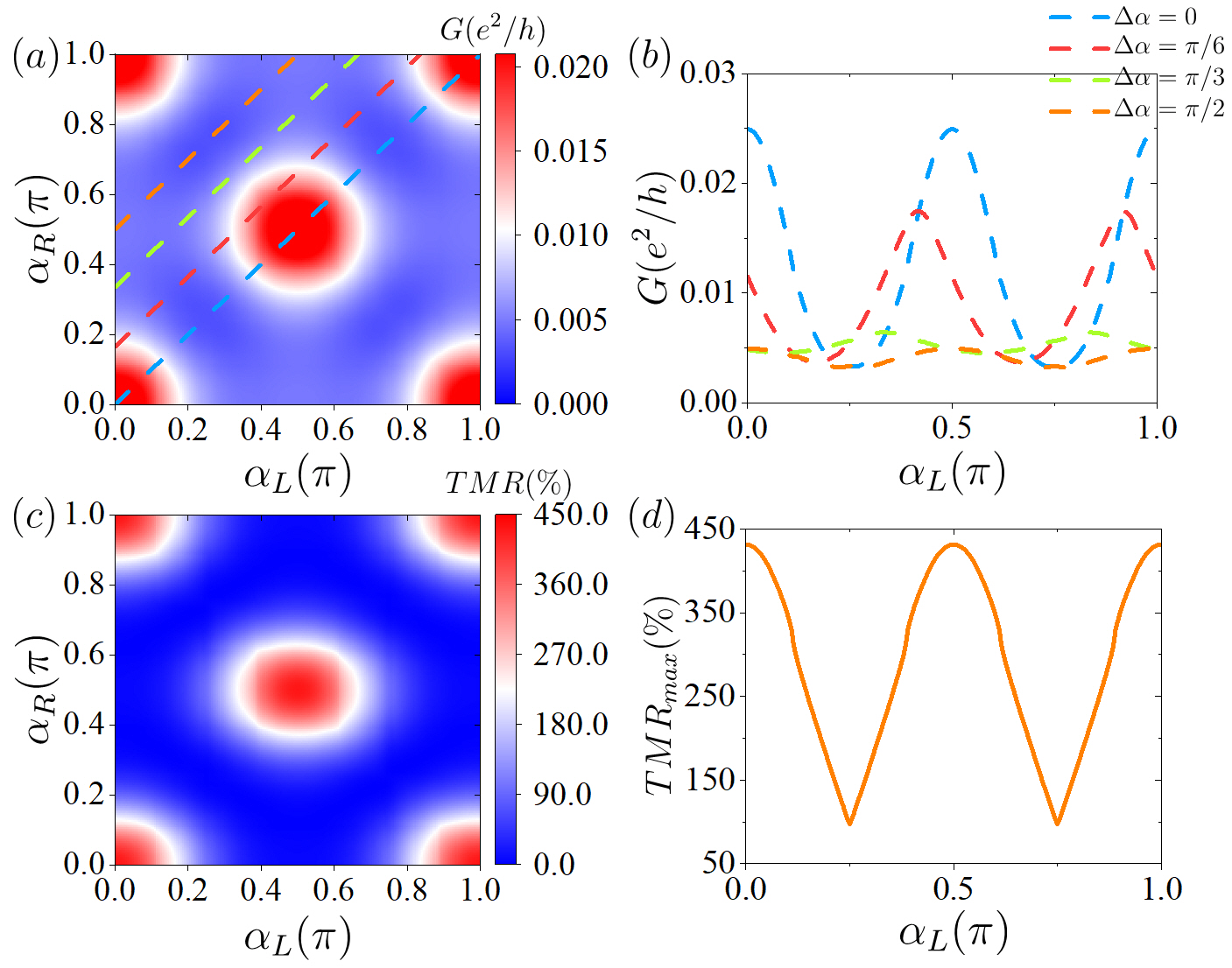}}
    \caption{
(a) The conductance versus the orientation $\alpha_L$ and $\alpha_R$.
(b) The conductance versus $\alpha_L$ under different $\Delta \alpha$ ($\Delta \alpha=\alpha_R-\alpha_L$), which corresponds to colored dashed lines in (a), respectively. The data are extended beyond the original range to $[0, \pi]$.
(c) The TMR ratio versus the orientation $\alpha_L$ and $\alpha_R$.
Note that for a fixed $\alpha_L$, we rotate $\alpha_R$ and calculate each corresponding conductance, denoted as $G_2$, and we choose the minimum value among them as $G_1$, then TMR ratio can be obtained from Eq.(\ref{TMR}). (d) The maximum TMR ratio for each fixed $\alpha_L$ when $\alpha_R$ rotates from 0 to $\pi$.}
    \label{Fig4}
\end{figure}

In Fig. \ref{Fig4}(a), we show the conductance versus the orientation $\alpha_L$ and $\alpha_R$.
The conductance $G$ has large values while $(\alpha_L,\alpha_R)=(0/\pi,0/\pi)$
and $(0.5\pi, 0.5\pi)$, because in these $(\alpha_L,\alpha_R)$, the dominant spin in the left AM, the dominant spin in the right AM, and the transport direction (x-direction) are perfectly matching.
Deviations from these values of $(\alpha_L,\alpha_R)$ result in a sharp reduction in conductance.
For example, when $(\alpha_L,\alpha_R)=(\pi/4,\pi/4)$, the conductance is significantly suppressed [see Fig. \ref{Fig4}(a)], because the orientations of the left and right AMs, although aligned with each other, do not match the transport direction ($x$-axis).

For any fixed $\alpha_L$, the conductance versus $\alpha_R$ has the same symmetry,
$G(\alpha_L, \alpha_R)=G(\alpha_L, \pi-\alpha_R)$, as that in Fig. \ref{Fig3}(a).
Similarly, considering the equivalency between the right AM and the left AM, the conductance $G(\alpha_L,\alpha_R)=G(\pi-\alpha_L,\alpha_R)$ is also satisfied. Additionally, the conductance is also centrosymmetric with respect to the orientation $\left(\alpha_L, \alpha_R \right)=\left( \frac{\pi}{2}, \frac{\pi}{2} \right) $, satisfying the relation: $G(\alpha_L, \alpha_R)=G(\pi-\alpha_L, \pi-\alpha_R)$.

Due to the current conservation, the linear conductance from the left AM to the right AM must be equal to the conductance from the right AM to the left AM, that is $G(\alpha_L, \alpha_R)=G(\alpha_R, \alpha_L)$, and Fig. \ref{Fig4}(a) is symmetric about the diagonal line with $\alpha_R=\alpha_L$. Therefore, simultaneously considering the joint symmetry operation and the current conservation, the conductance should satisfy $G(\alpha_L, \alpha_R)= G(\pi-\alpha_R, \pi-\alpha_L)$, indicating that Fig. \ref{Fig4}(a) is also symmetric about the negative-slope diagonal line with $\alpha_R=\pi-\alpha_L$.

In Fig. \ref{Fig4}(b), we plot the distribution of conductances versus $\alpha_L$ along four colored dashed lines for different values of $\Delta \alpha=\alpha_R-\alpha_L$. Obviously, even though the relative AM angle $\Delta \alpha$ is always unchanged, the conductance still varies as $\alpha_L$ rotates. This is attributed to the highly anisotropic Fermi surface of the AM.
As for the blue curve with $\Delta \alpha=0$, the orientations of the left and right AMs are the same and should, in principle, facilitate the electron tunneling and result in the highest conductance. However, as $\alpha_{L}=\alpha_{R}$ rotates, the conductance oscillates both maximum and minimum values at certain different points.
Specifically, there is a period of $0.5\pi$, since the band structure returns to its original form just with the reversed spin states after $0.5\pi$ rotation.
The behaviors for the other dashed lines with the non-zero $\Delta \alpha$ are similar, except that the positions of maximum and minimum points shift.

In Fig. \ref{Fig4}(c), we plot the TMR ratio versus the orientation $\alpha_L$ and $\alpha_R$. In detail, for each fixed $\alpha_L$, the conductance $G$ will change as  $\alpha_R$ varies from 0 to $\pi$, which is denoted as $G_{2}$ [$G_{2}=G(\alpha_L,\alpha_R)$]. And the minimum conductance value among them is just chosen as $G_{1}$, i.e. $G_{1}(\alpha_L)\equiv \min[G(\alpha_L,\alpha_{R}\in(0,\pi))]$.
Similar to Fig.~\ref{Fig4}(a), the TMR ratio has the large value at $(\alpha_L,\alpha_R)=(0/\pi,0/\pi)$ and $(0.5\pi, 0.5\pi)$, and deviations from these values of $(\alpha_L,\alpha_R)$, it sharp reduces.
In addition, the TMR ratio also satisfies the relation  $TMR(\alpha_L, \alpha_R)=TMR(\pi-\alpha_L, \alpha_R)$ and $TMR(\alpha_L, \alpha_R)=TMR(\alpha_L, \pi-\alpha_R)$. Considering the combined effect above, the relation $TMR(\alpha_L, \alpha_R)=TMR(\pi-\alpha_L, \pi-\alpha_R)$ is also obtained.

From the results in Fig. \ref{Fig4}(c), we can extract the maximum TMR ratios for each fixed $\alpha_L$, as shown in Fig. \ref{Fig4}(d). We can find at $\alpha_{L}=0$ or $0.5\pi$, the AM-I-AM junctions possess a higher symmetry with respect to the transport direction, and the change of the conductance $G$ becomes more sensitive as $\alpha_{R}$ varies, which results in high TMR ratios.

\section{\label{sec4} The dependence of TMR effect on the orientation of spin}

In the above section, with the fixed orientations of spins,
we have discussed how the orientations of AMs affect the conductance and TMR ratio.
In this section, we will study how the orientations of spins ${\bf s}_{L/R}$
affect the conductance and TMR ratio with the fixed orientations of AMs $\alpha_L=\alpha_R=0$.

Due to the anisotropy of the Fermi surface, the conductance will vary with the rotation of the orientation of AM even for a constant $\Delta \alpha$.
However, since spin is an intrinsic degree of freedom, things will become quite different.
To investigate this, we fix the orientation of the spin $\mathbf{s}_L$ in left AM at a random angle $(\theta_L, \phi_L)$, see Fig. \ref{Fig1}(c). For the spin $\mathbf{s}_R$ in right AM, we can first define its orientation at $(\theta_R, \phi_R)=(\theta_L-\Delta \theta, \phi_L)$, where $\Delta \theta$ is the relative spin angle between the left and right spins, and then rotate it around the orientation $\mathbf{s}_L$ with this fixed relative spin angle $\Delta \theta$ and the rotational phase angle $\beta$, see Fig. \ref{Fig1}(d).

\begin{figure}[!htb]
\centerline{\includegraphics[width=\columnwidth]{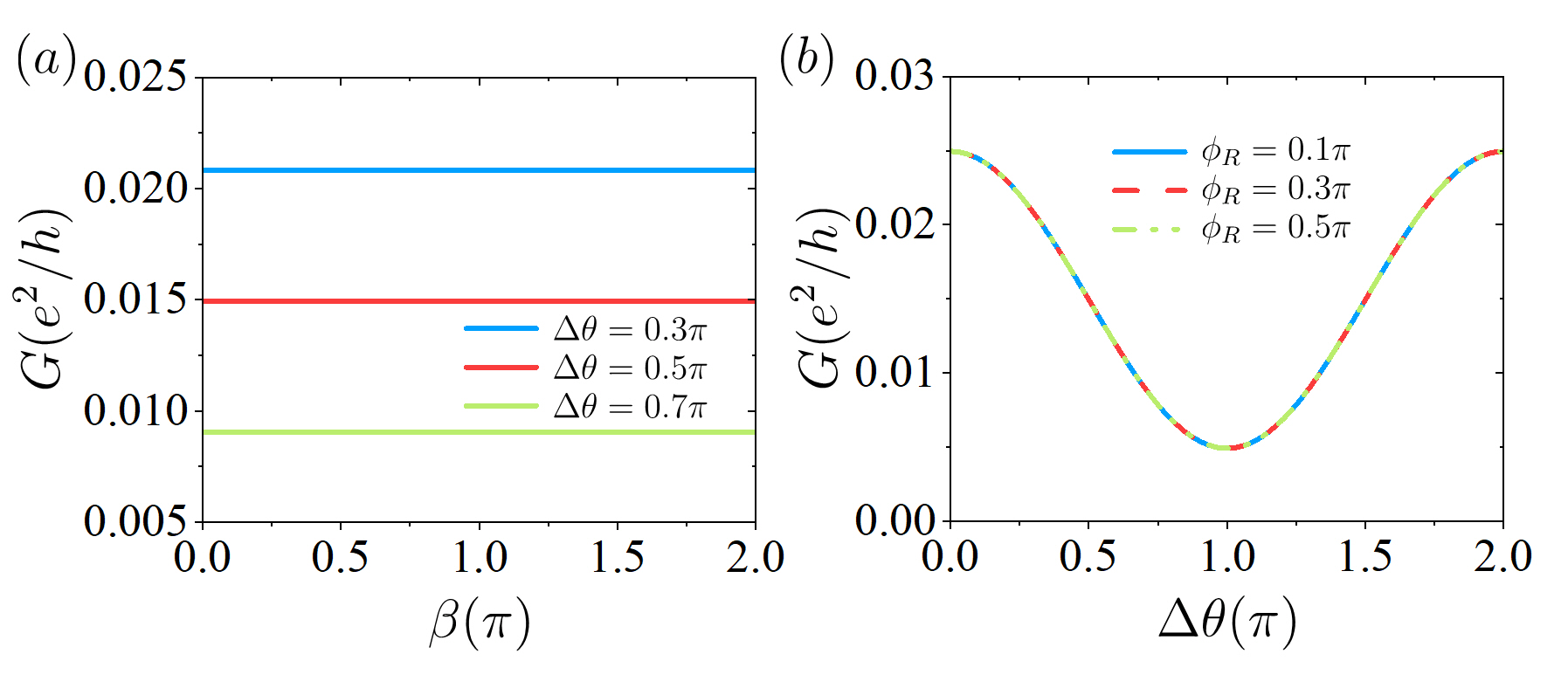}}
    \caption{
(a) The conductance versus the rotational phase angle $\beta$ under different fixed relative spin angle $\Delta \theta$. The orientation of the spin for the left AM is set at $(\theta_L, \phi_L)=(\frac{\pi}{3}, \frac{\pi}{4})$ randomly.
(b) The conductance versus the relative spin angle $\Delta \theta$ under different azimuthal angle $\phi_R$.
The orientation of the left and right AM is set at $\alpha_L=\alpha_R=0$.
 }
    \label{Fig5}
\end{figure}

In Fig. \ref{Fig5}(a), we plot the conductance $G$ versus the rotational phase angle $\beta$ under different fixed relative spin angle $\Delta \theta$.
It can be seen that, for a randomly fixed orientation of the left spin, as long as the relative spin angle between the left and right spins remains unchanged, the conductance of the system stays invariant. This can be explained by the spin rotation operator $U_{s}=e^{-i\frac{\beta}{2}\bm{\sigma}\mathbf{s}_L}$ by angle $\beta$ about the orientation of the left spin $\mathbf{s}_L$. Before and after the rotation, the Hamiltonian of the whole system can be linked by a unitary transformation: $H^{\prime}=U_s H U_s ^{\dagger}$. Since the unitary transformation will not affect the observable quantity, the conductance should remain unchanged.

In Fig. \ref{Fig5}(b), we also show the conductance versus the relative spin angle $\Delta \theta$ under different azimuthal angles $\phi_R$.
The orientation of the spin of left AM is fixed at $(\theta_L, \phi_L)=(0,0)$, and the $\theta_R$  of the spin of right AM rotates from 0 to $2\pi$ for different $\phi_R$. Since the conductance only depends on the relative spin angle $\Delta \theta$ between the two spins, no matter at what azimuthal $\phi_R$, the conductance remains the same.
While the relative spin angle $\Delta \theta$ increases from 0 to $2\pi$, the conductance will firstly decrease and then increase.
Roughly, the conductance can be determined by projecting the right spin wavefunction $\left |\mathbf{s}_R\right \rangle$ onto the left spin wavefunction $\left |\mathbf{s}_L\right \rangle$.
In detail, considering $s_L$ is fixed at  $(\theta_L, \phi_L)=(0,0)$ and $s_{R}$ is fixed at  $(\theta_R, \phi_R)=(\Delta \theta, \phi_R)$, the spin wavefunctions for them pointing to the positive direction are $\left | \mathbf{s}_L\right \rangle = \binom{1}{0} $ and  $\left | \mathbf{s}_R\right \rangle = \binom{\cos(\Delta \theta/2)}{\sin(\Delta \theta/2)e^{i\phi_{R}}} $.
This approximately gives $G\propto \left | \left \langle \mathbf{s}_L | \mathbf{s}_R \right \rangle \right |^2 = \cos^2 \frac{\Delta \theta}{2}$. The conductance is symmetric about $\Delta \theta=\pi$ where the value is the minimum, and the period is $2\pi$.

\section{\label{sec5} The combined effect of the orientations of AM and spin}

In Secs. \ref{sec3} and \ref{sec4}, we have discussed how the orientations of AM and spin affect the conductance and the TMR ratio.
In this section, we will combine these two aspects together and study their joint effect.

\begin{figure*}[t]
\centerline{\includegraphics[width=1.5\columnwidth]{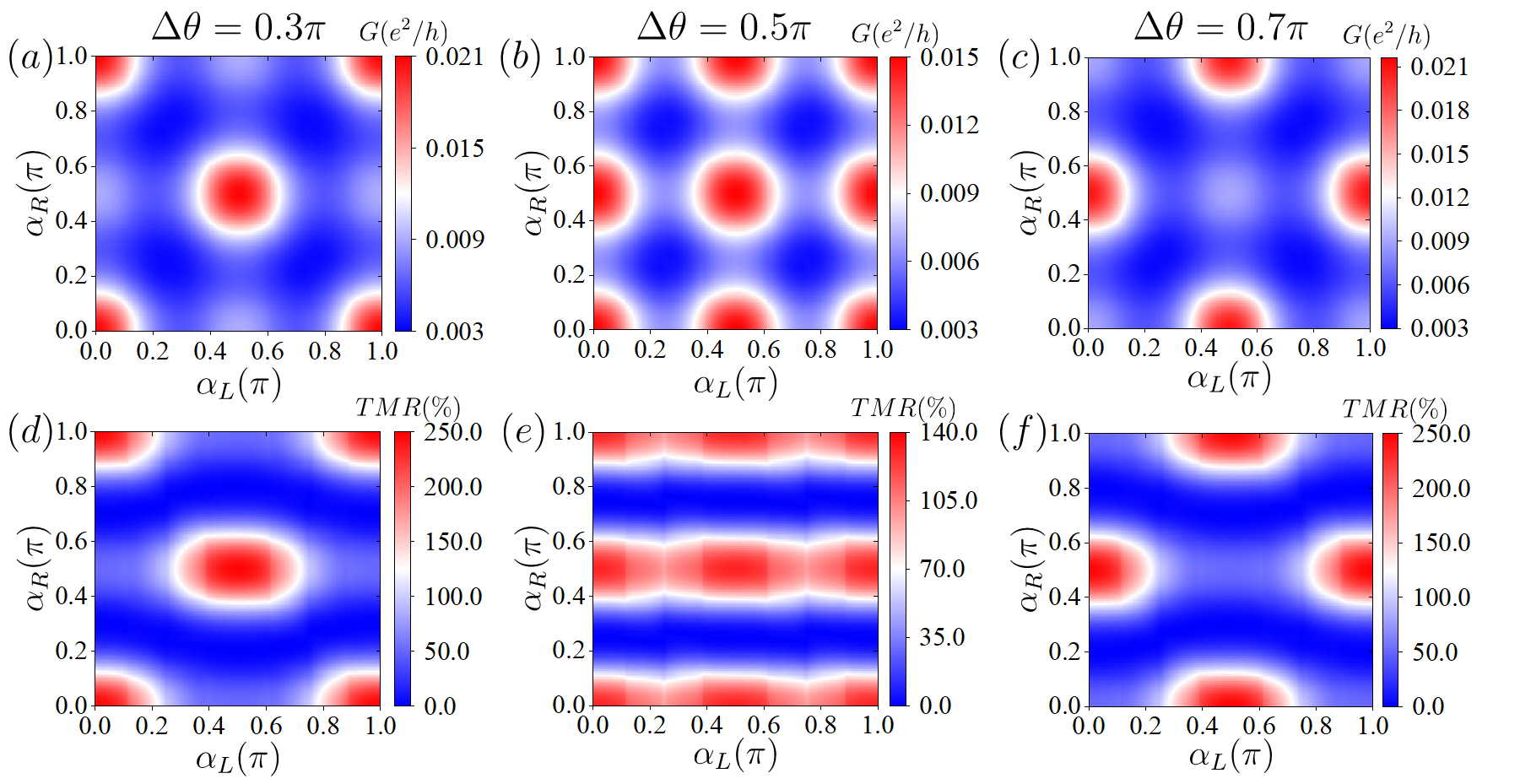}}
    \caption{The conductance and the TMR ratio versus the orientations of the left and right AMs, $\alpha_L$ and $\alpha_R$, under different relative spin angle $\Delta \theta$.
    (a-c) are the conductance and (d-f) are the corresponding TMR ratio.
    The relative spin angle $\Delta \theta = 0.3\pi$ in (a, d),
    $0.5\pi$ in (b, e), and $0.7\pi$ in (c, f).
}
\label{Fig6}
\end{figure*}

In Figs. \ref{Fig6}(a-c), we plot the conductance versus the orientations of the left and right AMs, $\alpha_L$ and $\alpha_R$, under different fixed relative spin angle $\Delta \theta$.
%These figures are similar to Fig. \ref{Fig4}(a) with $\Delta \theta=0$ and exhibit similar symmetry.
In Fig. \ref{Fig6}(a) with $\Delta \theta=0.3\pi$, the conductance at orientation $(\alpha_L, \alpha_R)=(0, 0.5\pi)$ is increased comparing to Fig. \ref{Fig4}(a).
This is because when $(\alpha_L, \alpha_R)=(0, 0.5\pi)$, the spin texture for the left and right AMs is no longer completely antiparallel as in the case of $\Delta \theta =0$. Actually, we can find these conductances for the center of edges tend to be local maxima (white regions). It is clearer in Fig. \ref{Fig6}(b) with $\Delta \theta=0.5\pi$, where the maximum conductance appears at the center, the four corners, and the centers of four edges. The explanation will be demonstrated later. While in Fig. \ref{Fig6}(c) with $\Delta \theta=0.7\pi$, we can find the maximum conductance changes its position from the corners and center in Fig. \ref{Fig6}(a) to the centers of four edges, considering $\mathbf{s}_{L}$ and $\mathbf{s}_R$ tend to be antiparallel when $\alpha_{L}=\alpha_R=0$.

When $\Delta \theta=0.5\pi$,  we can take $\alpha_L=0$ with $\alpha_R=0$ and $\alpha_R=0.5\pi$ as an example to discuss.
Since the conductance is only related to the relative spin angle $\Delta \theta$ but not to the specific orientation of spin, we can set the orientation of $\mathbf{s}_L$ along the $z$-axis at $(\theta_L, \phi_L)=(0,0)$.
The two left spin states corresponding to the red $(\left | 1 \right \rangle)$ and blue $(\left | 2 \right \rangle)$ elliptical energy surfaces in Fig. \ref{Fig1}(a) can be denoted as: $\left | 1 \right \rangle= \left | \uparrow  \right \rangle $ and $ \left | 2 \right \rangle =\left | \downarrow  \right \rangle $, respectively.
When $\alpha_R=0$ and the relative spin angle is $\Delta \theta=0.5\pi$, the orientation of $\mathbf{s}_R$ is in the $xy$ plane. For simplicity, we can denote these two spin states as $\left | 1^{\prime} \right \rangle=\left | \rightarrow \right \rangle = \frac{\left | \uparrow  \right \rangle + \left | \downarrow  \right \rangle }{\sqrt{2}}$ and $ \left | 2^{\prime} \right \rangle =\left | \leftarrow \right \rangle = \frac{\left | \uparrow  \right \rangle - \left | \downarrow  \right \rangle }{\sqrt{2}}$, respectively.
When the orientation of the right AM rotates to $\alpha_R=0.5\pi$, the two perpendicular elliptical energy surfaces exchange their positions, leading to a transformation in the two right spin states as $\left | 1^{\prime} \right \rangle=\left | \leftarrow \right \rangle$ and $\left | 2^{\prime} \right \rangle=\left | \rightarrow \right \rangle$.
Since the two spin states $\left | \leftarrow \right \rangle $ and $\left | \rightarrow \right \rangle $ only differs by a relative phase factor between $\left | \uparrow  \right \rangle $ and $\left | \downarrow  \right \rangle $, it should not affect transmission probability, as the transport properties depend mainly on the relative projection of the left and right spin states. Thus, the conductance is the same at $(\alpha_L, \alpha_R)=(0, 0)$ and $(\alpha_L, \alpha_R)=(0,0.5\pi)$. Furthermore, it is also easily obtainable that the conductance is equal at all nine values $(\alpha_L, \alpha_R)=(0/0.5\pi/\pi, 0/0.5\pi/\pi)$.

In Figs. \ref{Fig6}(d-f), we also plot the TMR ratio versus $\alpha_L$ and $\alpha_R$, corresponding to Figs. \ref{Fig6}(a-c), respectively.
When calculating the TMR ratio, the selection of the conductance $G_1$ and $G_2$ is consistent with that in Fig. \ref{Fig4}(c).
In Fig. \ref{Fig6}(d), the positions of the maximum TMR ratio are similar to those in \ref{Fig4}(c), since the relative spin angle $\Delta \theta$ is small, the orientation of spin in the left and right AM is approximately parallel.
While in Fig. \ref{Fig6}(e) with $\Delta \theta=0.5\pi$, the orientation of spin in the left and right AM is perpendicular.
The positions where the maximum value appears in this figure correspond to the positions in Fig. \ref{Fig6}(b), and the maximum values of $TMR(\alpha_L=0,\alpha_R)$ and $TMR(\alpha_L=0.5\pi,\alpha_R)$ are the same.
The overall magnitude of the TMR ratio decreases greatly compared to that of Fig. \ref{Fig4}(c), and the difference between the maximum TMR ratio with different $\alpha_L$ becomes less significant, as shown in the expansion of the red regions.
In Fig. \ref{Fig6}(f) with $\Delta \theta=0.7\pi$, the maximum TMR ratio shifts its position from the center and the four corners in Fig. \ref{Fig6}(d) to the centers of the four edges, since the orientation of spin in the left and right AM is close to antiparallel.

Moreover, since the orientation of AM and the orientation of spin are two independent degrees of freedom, rotating the fixed relative spin angle 
$\Delta\theta$ will not affect the symmetry features with respect to $\alpha$.
Therefore, the conductance exhibits the same symmetry relations as in Fig. \ref{Fig4}(a).
Due to the symmetry relations: $G(\alpha_L, \alpha_R)=G(\alpha_L, \pi-\alpha_R)$ and $G(\alpha_L, \alpha_R)=G(\pi-\alpha_L, \alpha_R)$, the TMR ratio also satisfies the relation: $TMR(\alpha_L, \alpha_R)=TMR(\pi-\alpha_L, \alpha_R)$ and $TMR(\alpha_L, \alpha_R)=TMR(\alpha_L, \pi-\alpha_R)$. Taking into account the combined effect above, the relation $TMR(\alpha_L, \alpha_R)=TMR(\pi-\alpha_L, \pi-\alpha_R)$ is obtained, which also exhibits the same symmetry relations as in Fig. \ref{Fig4}(c).
\\

\section{\label{sec6} Conclusion}

In conclusion, we theoretically propose a universal and concise altermagnetic sandwich device to investigate its exotic TMR effect and transport properties.
Based on the nonequilibrium Green's function method and the Landauer-B\"uttiker formula, we calculate the conductance and the TMR ratio in this AM-I-AM junction.
We demonstrate the emergence of the AM-based TMR effect, and illuminate the properties of the transport and TMR ratio by rotating various orientations of AM $(\alpha_L, \alpha_R)$ and the left spin $(\theta_L, \phi_L)$ and right spin $(\theta_R, \phi_R)$.
In summary, for the six above parameters, only three parameters are ultimately needed: $\alpha_L, \alpha_R$, and the relative spin angle between spins $\Delta \theta$.
Furthermore, symmetry relations between these orientation parameters are systematically analyzed.
This work thoroughly and comprehensively discusses how various magnetic configurations in the altermagnetic system affect the transport and the TMR ratio, highlighting the remarkable and distinctive features of the AM-based TMR effect.
It could open up horizons for the next-generation information technologies on the altermagnetic platform and promote the development of spintronics.

\section*{\label{sec6}ACKNOWLEDGMENTS}
This work was financially supported by
the National Key R and D Program of China (Grant No. 2024YFA1409002),
the National Natural Science Foundation of China (Grants No. 12374034 and No. 12447146), the Innovation Program for Quantum Science and Technology (Grant No. 2021ZD0302403), and the Postdoctoral Fellowship Program of CPSF under Grant Number GZB20240031. The computational resources are supported by the High-Performance Computing Platform of Peking University.	

\appendix
\section{\label{A} Discretization of the Hamiltonian}

\begin{widetext}
 The discrete Hamiltonians for left and right AMs $H_{L}$ and $H_{R}$:
\begin{align}
	H_{R} = &\sum_{\substack{i_{x} \geqslant  N_x+1  \\	1  \leqslant i_{y} \leqslant N_y }}
		\Psi_{i}^{R \dagger } \check{H}_{0}^{R} \Psi_{i}^{R} %\notag \\
		 + \left[
            \sum_{\substack{i_{x} \geqslant  N_x+1 \\ 	1  \leqslant i_{y} \leqslant  N_y }}
		\Psi_{i}^{R \dagger } \check{H}_{x}^{R} \Psi_{i+\delta x}^{R}
            +\sum_{\substack{i_{x} \geqslant  N_x+1\\ 1  \leqslant i_{y} \leqslant  N_y-1 }}
		\Psi_{i}^{R \dagger } \check{H}_{y}^{R} \Psi_{i+\delta y}^{R} \right. \notag \\
            & + \left.
            \sum_{\substack{i_{x} \geqslant  N_x+1\\ 1  \leqslant i_{y} \leqslant  N_y-1 }}
		\Psi_{i}^{R \dagger } \check{H}_{xy}^{R} \Psi_{i+\delta x + \delta y}^{R}
            + \sum_{\substack{i_{x} \geqslant  N_x+1\\ 2  \leqslant i_{y} \leqslant  N_y }}
		\Psi_{i}^{R \dagger } \check{H}_{x\bar{y}}^{R} \Psi_{i+\delta x - \delta y}^{R}
            +\text { H.c. }\right],
\end{align}
\begin{align}
H_{L} = & \sum_{\substack{i_{x} \leqslant 0 \\ 1 \leqslant i_{y} \leqslant N_y }}
\Psi_{i}^{L \dagger } \check{H}_{0}^{L} \Psi_{i}^{L} %\notag \\
 + \left[
    \sum_{\substack{i_{x} \leqslant -1 \\ 1 \leqslant i_{y} \leqslant N_y }}
    \Psi_{i}^{L \dagger } \check{H}_{x}^{L} \Psi_{i+\delta x}^{L}
    + \sum_{\substack{i_{x} \leqslant 0 \\ 1 \leqslant i_{y} \leqslant N_y-1 }}
    \Psi_{i}^{L \dagger } \check{H}_{y}^{L} \Psi_{i+\delta y}^{L} \right. \notag \\
& + \left.
    \sum_{\substack{i_{x} \leqslant -1 \\ 1 \leqslant i_{y} \leqslant N_y-1 }}
    \Psi_{i}^{L \dagger } \check{H}_{xy}^{L} \Psi_{i+\delta x +\delta y}^{L}
    + \sum_{\substack{i_{x} \leqslant -1 \\ 2 \leqslant i_{y} \leqslant N_y }}
    \Psi_{i}^{L \dagger } \check{H}_{x\bar{y}}^{L} \Psi_{i+\delta x - \delta y}^{L}
    + \text{H.c.} \right].
\end{align}
Here, $\Psi_{i}^{L(R)}=(\Psi_{L(R)i \uparrow}, \Psi_{L(R)i \downarrow})^{T}$ in which $\Psi_{L(R)i \uparrow (\downarrow)}$ is the annihilation operator of electron with spin up (down) on lattice site $i=(i_x, i_y)$ in the left (right) AM.
The subscripts $i+\delta x$ ($i+\delta y)$ represents the nearest-neighbor hopping from the $i$th site along the $x$ ($y$) direction, while the subscripts $i+\delta x+\delta y$ ($i+\delta x-\delta y)$ represents the next-nearest-neighbor hopping from the $i$th site, as shown in Fig. \ref{Fig1}(b).
The Hamiltonian matrices are:
\begin{align}
\check{H}_{0}^{L(R)} &=\left(\frac{4t_0}{a^2}-\mu_{L(R)} \right) \sigma_0, \\
\check{H}_{x}^{L(R)} &=-\frac{t_0}{a^2}\sigma_0-\frac{t_J\cos2\alpha_{L(R)}}{a^2} \bm{\sigma} \cdot \mathbf{s}_{L(R)}, \\
\check{H}_{y}^{L(R)} &=-\frac{t_0}{a^2}\sigma_0+\frac{t_J\cos2\alpha_{L(R)}}{a^2} \bm{\sigma} \cdot \mathbf{s}_{L(R)},
\end{align}
and
\begin{align}
\check{H}_{xy}^{L(R)} &=-\frac{t_J\sin2\alpha_{L(R)}}{2a^2} \bm{\sigma} \cdot \mathbf{s}_{L(R)},
\\
\check{H}_{x\bar{y}}^{L(R)} &=-\check{H}_{xy}^{L(R)}.
\end{align}

\begin{figure*}[htbp]
\centerline{\includegraphics[width=0.75\columnwidth]{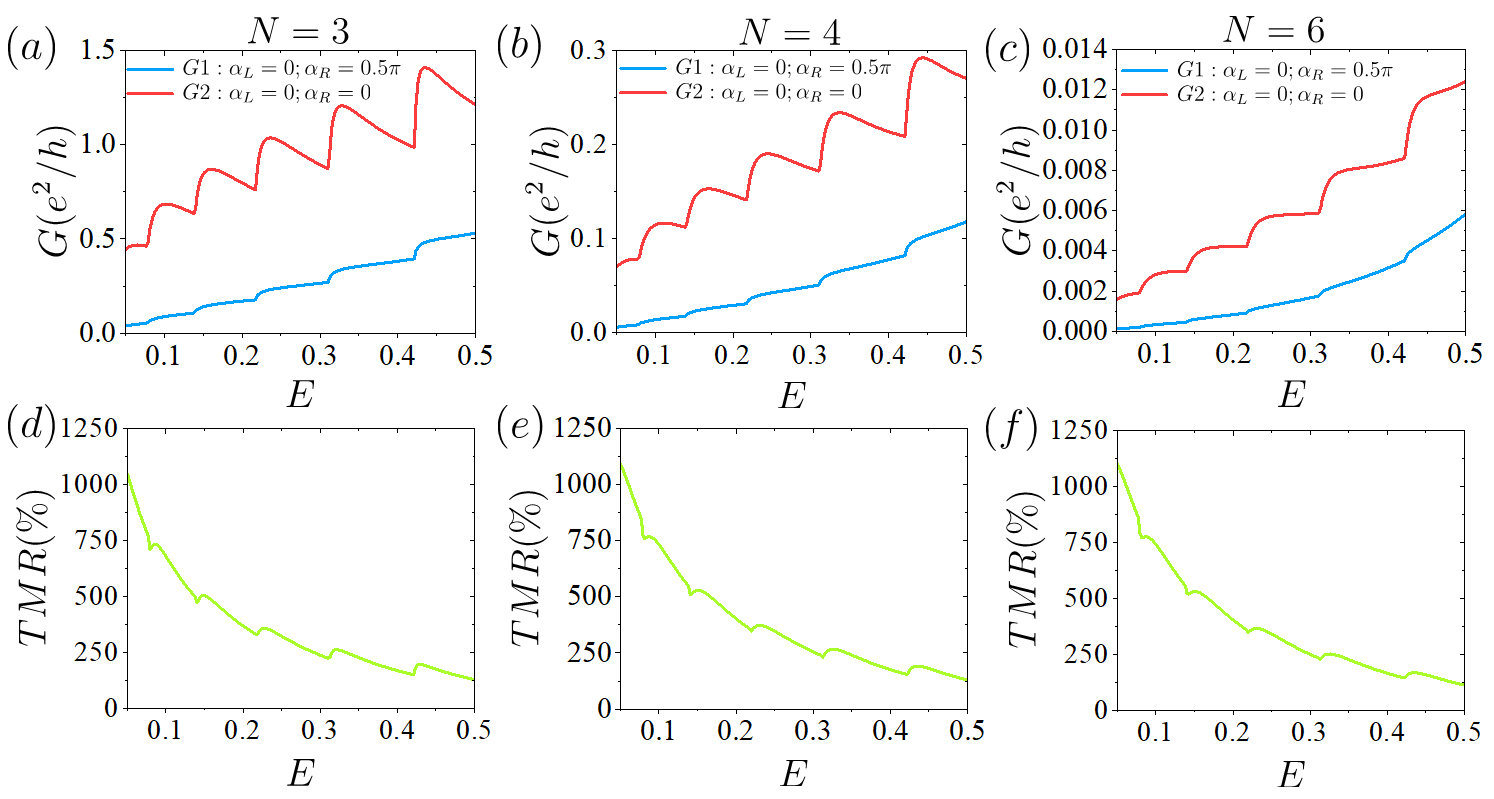}}
    \caption{
The conductance $G$ (a-c) and the TMR ratio (d-f) versus the Fermi energy $E$ under different lengths of the center insulating layer $N_x=3$ in (a,d), $N_x=4$ in (b,e), and $N_x=6$ in (c,f).
$t_j=0.5$ and the other unmentioned parameters are consistent with those in Fig. \ref{Fig2}(c-d).
 }
    \label{R1}
    \end{figure*}

The discrete Hamiltonian for the center insulator $H_I$ is:
\begin{align}
	H_{I} = &\sum_{\substack{1  \leqslant i_{x} \leqslant  N_x  \\	1  \leqslant i_{y} \leqslant N_y }}
		\Psi_{i}^{I \dagger } \check{H}_{0}^{I} \Psi_{i}^{I}
        + \left[
            \sum_{\substack{1  \leqslant i_{x} \leqslant  N_x-1 \\ 	1  \leqslant i_{y} \leqslant  N_y }}
		\Psi_{i}^{I \dagger } \check{H}_{x}^{I} \Psi_{i+\delta x}^{I}
            +\sum_{\substack{1  \leqslant i_{x} \leqslant  N_x\\ 1  \leqslant i_{y} \leqslant  N_y-1 }}
		\Psi_{i}^{I \dagger } \check{H}_{y}^{I} \Psi_{i+\delta y}^{I}      +\text { H.c. }\right],
\end{align}
Here, $\Psi_{i}^{I}=(\Psi_{Ii \uparrow}, \Psi_{Ii \downarrow})^{T}$ in which $\Psi_{Ii \uparrow (\downarrow)}$ is the annihilation operator of electron with spin up (down) on lattice site $i=(i_x, i_y)$ in the center insulator. The Hamiltonian matrices are
\begin{align}
\check{H}_{0}^{I} &=\left(\frac{4t_0}{a^2}-\mu_{I} \right) \sigma_0, \\
\check{H}_{x}^{I} &=-\frac{t_0}{a^2}\sigma_0, \\
\check{H}_{y}^{I} &=-\frac{t_0}{a^2}\sigma_0.
\end{align}

The tunneling Hamiltonian between different regions is written as:
	\begin{equation}
		\begin{aligned}
			H_{T} & = \sum_{\substack{i_{x} = 0 \\ 1 \leqslant i_{y} \leqslant N_y }}
    \Psi_{i}^{L \dagger } \check{T} \Psi_{i+\delta x}^{I}
    + \sum_{\substack{i_{x} = N_x \\ 1 \leqslant i_{y} \leqslant N_y }}
    \Psi_{i}^{I \dagger } \check{T} \Psi_{i+\delta x}^{R}
    + \text{H.c.},
		\end{aligned}
	\end{equation}
with the hopping matrix $\check{T}=\mathrm{diag}(t,t)$. $t=-\frac{t_0}{a^2}$ is the coupling energy between the insulator and AMs.
\end{widetext}

\section{\label{B} The influence of the length of the insulating layer}

Quantum tunneling effects can be introduced in the system by introducing an insulating layer in the intermediate region. Moreover, this insulating layer can also facilitate the independent tuning of the orientations of the AM and spin at both ends, and significantly improve the sensitivity and reduce the power consumption in practical applications \cite{AN2025116174,Edelstein_2007,Magnetic_Tunnel_2014}.
In this part, we explore the influence of the length of the center insulating layer $N_x$ on the numerically calculated results.

In Fig. \ref{R1}, we plot the conductance and the TMR ratio versus the Fermi energy under different lengths of the insulating layer $N_x$. We select several sets of the layer with $N_x=3, 4, 5, 6$, where $N=5$ corresponds to the case shown in Fig. \ref{Fig2}. The other calculation parameters are consistent with those in Fig. \ref{Fig2}.
It is obvious that as the insulating layer becomes thicker, the tunneling of the electrons is suppressed, leading to a decrease in the conductance.
However, we can find that the TMR ratio almost remains constant. This is because the increase in the length of the insulating layer mainly raises an isotropic potential barrier in the process of spin tunneling, which will not affect the spin polarization of the left and right AMs. These results reflect the robustness of TMR in the AM-I-AM junction.

%\section*{REFERENCES}
\bibliography{reference.bib}

\end{document}